\documentclass[prd,amssymb,floatfix,superscriptaddress,nofootinbib]{revtex4}
\usepackage{amsmath,amssymb,color}
\usepackage{graphicx}

\newcommand{\be}{\begin{eqnarray}}
\newcommand{\ee}{\end{eqnarray}}

\def\ben{\begin{equation}}
\def\een{\end{equation}}
\def\bena{\begin{eqnarray}}
\def\eena{\end{eqnarray}}

\begin{document}

\title{On the many saddle points description of quantum black holes}

\author{Cristiano Germani}
\email{cristiano.germani@physik.uni-muenchen.de}
\affiliation{ Arnold Sommerfeld Center, Ludwig-Maximilians-Universit\"at, Theresienstr. 37, 80333 Muenchen, Germany}

\begin{abstract}
Considering two dimensional gravity coupled to a CFT, we show that a semiclassical black hole can be described in terms of two Liouville theories matched at the horizon. The black hole exterior corresponds to a space-like while the interior to a time-like Liouville theory.  This matching automatically implies that a semiclassical black hole has an infinite entropy. 
The path integral description of the time-like Liouville theory (the Black Hole interior) is studied and it is found that the correlation functions of the coupled CFT-gravity system are dominated by two (complex) saddle points, even in the semiclassical limit. We argue that this system can be interpreted as two interacting Bose-Einstein condensates constructed out of two degenerate quantum states. In AdS/CFT context, the same system is mapped into two interacting strings intersecting inside a three-dimensional BTZ black hole. Finally, we discuss why, beyond the semiclassical approximation, we expect no firewalls appearing in our system.
\end{abstract}

\maketitle

\section{Introduction}

In General Relativity (GR) black holes (BH) are stable geometrical spaces equipped by an event horizon. Event horizons (sometimes called Cauchy horizons) are light-like closed surfaces with the property that the interior cannot causally influence the exterior. 

In any consistent theory of gravity, a BH is inexorably formed whenever collapsing matter enters inside a hoop of horizon size \cite{hoop,sarah}. After this, the whole information regarding the pre-BH physics, if not lost, must be stored inside the horizon, as the BH exterior carries ``no hair" \cite{nohair}.  

Black holes are also endowed by a spacetime curvature singularity in their center. There, curvatures become so large that quantum physics must take over.  This same fact, would imply that the path integral of gravity cannot possibly be dominated by the general relativistic (classical) BH solution, at least very close to the singularity. 
 
Does this implies that black holes can be treated classically far away from the singularity? The answer seems to be negative, even close to the horizon where curvatures are much smaller than the Planck scale $M_p$.

Indeed, a long time ago, Hawking has shown that black holes evaporate quantum mechanically with a black body spectrum of temperature $T$, which depends only (in the uncharged non-rotating case) upon the BH mass \cite{how}. Just after this important discovery, Hawking himself also realized that this (thermal) evaporation, would completely destroy the information stored inside the horizon.

In modern language the information loss can be stated as following:

Let us consider, for simplicity, GR minimally coupled to a massless scalar field. The quantum partition function of this system is
\be\label{path}
Z[g;\phi]=\int {\cal D}g{\cal D}\phi e^{\frac{i}{\hbar}S[g;\phi]}\ ,
\ee
where $g_{\alpha\beta}$ is the spacetime metric, $\phi$ the massless scalar field and the action
\be
S[g;\phi]=-\frac{1}{2}\int d^4x \sqrt{-g}\left(M_p^2 R+\partial_\alpha\phi\partial^\alpha\phi\right)\ .
\ee
Given the relevant boundary conditions for the path integral (\ref{path}), and considering a mass $M\gg M_p$, there is only one (real) saddle point of this action, i.e. a large GR black hole solution. A large BH means that the horizon curvature is far below the Planck scale so that perturbative quantum gravity is a good approximation far enough from the singularity. One can then consider quantum fluctuations of both the metric and the scalar around the BH geometry. Quantum fluctuations with energy much smaller than $M_p$ are dominated by the scalar. Thus, with a good approximation, we can neglect quantum gravity contributions and just consider scalar fluctuations, i.e.
\be
\langle\phi(t,x)\phi(t',x')\rangle\simeq \int_{g_{\rm BH}}{\cal D}\phi\ \phi(t,x)\phi(t',x')e^{\frac{i}{\hbar}S[g_{\rm BH};\phi]}\ ,
\ee 
where $g_{\rm BH}$ is the classical BH metric.

In this case, it has been shown in \cite{barbon} that, outside the horizon, the two point correlation functions of $\phi$ for $\Delta t=t-t'\gg \beta$ are
\be
\langle \phi(t,x)\phi(t',x)\rangle\sim e^{-\frac{\Delta t}{\beta}}\ ,
\ee
where $\beta$ is the inverse temperature of the (evaporating) quantum black hole.

In other words, the correlation functions of the scalar decay exponentially in time, so that ``information" is lost after a large time interval ($\gg \beta$).

A partial resolution to this puzzle was given by Maldacena in the context of three-dimensional asymptotically Anti DeSitter (AdS) spacetimes where a BH can stay forever in thermal equilibrium \cite{malda}. Maldacena realized that the same boundary conditions for the integration over topologies in the path integral \eqref{path}, have (at least) a second (complex) saddle point corresponding to thermal AdS (TAdS). Thus in calculating correlation functions one should at least consider, after Wick rotating time,
\be
\langle\phi(t,x)\phi(t',x')\rangle\simeq \int_{g_{\rm BH}}{\cal D}\phi\ \phi(t,x)\phi(t',x')e^{-\frac{S[g_{\rm BH};\phi]}{\hbar}}+\int_{g_{{\rm TAdS}}}{\cal D}\phi\ \phi(t,x)\phi(t',x')e^{-\frac{S[g_{{\rm TAdS}};\phi]}{\hbar}}\ ,\nonumber
\ee 
where $g_{\rm TAdS}$ is the thermal AdS metric.

Maldecena then showed that while the correlation functions calculated on the BH saddle point would decay in time the TAdS correlations would not.
In other words, the thermal AdS solution, obtained by Wick rotating time $t=i\tau$ and then fixing the periodicity to be $\tau\equiv\tau+\beta$, would generate a dominant contribution to the two-point correlation function at large time interval. However, although this configuration would restore partially information, in \cite{barbon} it has been shown that the only contribution to the path integral coming from TAdS would not be enough to restore the full information of the pre-BH physics.

Mainly, the failure of obtaining back the full information is due to the fact that the BH "weight" is exponentially larger than the TAdS by an entropy factor, i.e. for $\Delta t=t-t'\ll \beta$
\be\label{suppress}
\frac{\langle\phi(t,x)\phi(t',x)\rangle_{\rm TAdS}}{\langle\phi(t,x)\phi(t',x)\rangle_{\rm BH}}\sim e^{-{\cal S}_{\rm BH}}\ ,
\ee
where ${\cal S}_{\rm BH}\gg 1$ is the Bekenstein-Hawking entropy.

In a subsequent work \cite{lasthaw}, Hawking conjectured that the sum over all possible topologically trivial euclidean saddle points in the gravity path integral should restore information. Whereas, all saddle points with non-trivial topologies (such as BHs) would produce exponentially decaying (in time) correlations functions. However, still the question of whether non-exponentially suppressed (as in \eqref{suppress}) saddle points exist, has found no answer to date.  

More recently Dvali and Gomez, in a series of papers \cite{dvali}, put forward the idea that a black hole is a Bose-Einstein condensate of (weakly coupled) gravitons at a quantum critical point. The quantum critical point is defined by some parameters (for example the radius and mass of a star), such that whenever these are tuned to certain values (for example whenever the star radius is squeezed to the Schwarzschild radius) more quantum states suddenly coexist with same energy (for a toy model of this transition see \cite{alex}). There, therefore, correlations functions must be calculated by considering all the degenerate states equally. 

This sounds pretty much like the Maldacena (and Hawking) suggestion, where, contrary to the Maldacena case, it has been argued in \cite{dvali} that those co-existing quantum phases are {\it not} exponentially suppressed with respect to the classical BH solution. 

Another interesting fact comes from noticing that the critical point, describing a BH, is also unstable. The instability is such that gravitons are continuously lost so to decrease the mass and such that the black hole move from one critical point to another. For large black holes, the number of graviton lost is far smaller than the number present in the condensate and therefore an almost thermal spectrum is generated (Hawking radiation). This scenario might also unveil the nature of the Bekenstein black hole entropy \cite{bek}. Here, the entropy would just be the counting of the Bogoliubov modes, i.e. of the quantum fluctuations on the condensate. The number of Bogoliubov modes, and in turn the magnitude of the Bekenstein entropy, are however inversely proportional to $\hbar$ so that, in the limit $\hbar\rightarrow 0$ their number diverges. This conclusion implies that the Bekenstein entropy must be of pure quantum origin. Finally, in this scenario, the black hole singularity is also replaced by the condensate and so, the black hole interior is very far from being classical. 

All these ideas, clearly suggest that the information loss for "black holes" is due to the wrong assumption of considering, in the gravity path integral, only the classical (real) saddle point, i.e. the BH geometry. 

Here, we indeed show that a two-dimensional semiclassical BH is described as a statistical system where two (quantum) states co-exist. 

Note that we will use the name BH to mean the {\it semiclassical} system probed by an observer inside the region delimited by a horizon. However, we ask the reader not to be mislead by this. A semiclassical BH, as probed by a semiclassical observer, will be very far from the ``classical" BH solution. 

Note in addition that, in our specific case, the BH solution will be already found at one loop level, in a gravity-CFT system (that is why we quoted ``classical"). Thus, for us, {\it semiclassical} will really mean beyond the one loop approximation of the gravity matter system.

\section{Semiclassical two-dimensional black hole}

Let us consider a two dimensional theory of gravity coupled to a conformal field theory (CFT). We will follow the set up of \cite{norma}. The action of this system is
\be\label{action}
S=\frac{1}{2}\int d^2 x \sqrt{-g}\left(R-2\Lambda\right)+S_{CFT}\ .
\ee
The two dimensional metric can be gauge fixed in the light-cone coordinates $(\xi^+,\xi^-)$ to be 
\be\label{conf} 
ds^2=-e^{2\rho}d\xi^+ d\xi^-\ ,
\ee
where $\rho$ is a function of the light-cone coordinates. 

Standard text book calculations \cite{birrel} show that the stress tensor of the CFT, at one loop (i.e. in the ``semiclassical" limit), is restricted to satisfy the following equations for a given state $|\Psi\rangle$
\be\label{sys}
\langle T_{\pm\pm}\rangle&=&-\frac{\hbar\gamma}{12\pi}e^{\rho}\partial^2_\pm e^{-\rho}+t_{\pm}(\xi^\pm)\cr
\langle T_{\mu\nu}\rangle&=&\Lambda g_{\mu\nu}\cr
\langle T\rangle&=&\hbar\frac{\gamma}{24\pi}R\ ,
\ee 
where $\langle O\rangle\equiv\langle\Psi | O|\Psi\rangle$ and $t_{\pm}=\langle:T_{\pm\pm}:\rangle$ (the normal ordering of the energy-momentum tensor of the CFT) characterizes the state $|\Psi\rangle$. The factor $\gamma$ is related to the central charge of the field theory and counts, with appropriate sign, the total number of degrees of freedom. Explicitly, in the metric (\ref{conf}) with $e^{2\rho}>0$, we have that $\gamma=1-n_0-n_{1/2}$ \cite{kallosh} where the $1$ corresponds to the degree of freedom of the two dimensional ``graviton" and $n_i$ are the number of degrees of freedom of the CFT. Finally, the last of the equations (\ref{sys}) is called Liouville equation.

The system (\ref{sys}) defines spacetimes with constant curvature $R=\frac{48\pi\Lambda}{\gamma}$. We will seek for asymptotically Anti-DeSitter spaces (i.e. with constant negative curvature), therefore we shall take $\frac{\Lambda}{\gamma}<0$. This choice is very useful to be able to re-describe our system holographically via $AdS_3/CFT_2$ as in \cite{procopio}. At the same time, it considerably simplifies the system as the BH can be in thermal equilibrium, at least semiclassically. Finally, we will define $-\frac{24\pi\Lambda}{\gamma}\equiv \ell^{-2}$. 

Let us fix the vacuum to the Hartle-Hawking's \cite{hartle}. This vacuum has the important property that the propagator of a free field in a BH spacetime can be obtained as an analytic continuation of the Euclidean Green function and that the n-point correlations functions of interacting fields correspond, in Lorentzian signature, to thermal correlation functions \cite{gibbons}. The Hartle-Hawking vacuum is obtained by fixing $t_\pm(\xi^\pm)=\frac{\hbar\gamma}{48\pi}q=\mbox{const.}$ \cite{procopio}. We will also assume $q$ to be positive in order to obtain real solutions of the system (\ref{sys}). Moreover, a posterior, one can show that, for a black hole solution of the system (\ref{sys}),  the BH temperature is positively defined only for $q>0$ \cite{procopio}\footnote{Note the different sign conventions used in \cite{procopio}.}.

There are two solutions for the system (\ref{sys}) in the Hartle-Hawking vacuum, representing the choice of the ``time" direction
\be\label{sol1}
e^{2\rho}=\left\{ \begin{array}{ll} \frac{\hbar q\ell^2}{\sinh^2[\frac{1}{2}(\xi^+-\xi^-)\sqrt{q}]} &\mbox{for $(\xi^+-\xi^-)$ space-like} \\  -\frac{\hbar q\ell^2}{\cosh^2[\frac{1}{2}(\xi^+-\xi^-)\sqrt{q}]}& \mbox{for $(\xi^+-\xi^-)$ time-like}\end{array}\right.\ .
\ee
As we look for a static (or more precisely in thermal equilibrium) BH we will consider, outside the BH, the choice $(\xi^+-\xi^-)$ to be space-like. By defining $z\equiv\frac{\xi^+-\xi^-}{2}$ and $t\equiv\frac{\xi^++\xi^-}{2}$ as ``time", we have that the real solution of the Liouville equation is for
\be\label{sol}
e^{2\rho_{\rm out}}=\frac{\hbar q\ell^2}{\sinh^2[z\sqrt{q}]} \ ,
\ee
where $\hbar$ has been kept to explicitly show that this metric is non-trivial only quantum mechanically. However, from now on, we will fix units such that $\hbar=1$. The solution (\ref{sol}) follows from the Liouville equation
\be\label{Lout}
\tilde\square\rho_{\rm out}=\frac{1}{\ell^2}\ e^{2\rho_{\rm out}}\ ,
\ee
where the auxiliary metric $d\tilde s^2=-dt^2+dz^2$ has been used.
 
We can now find a set of coordinates such that the conformal metric (\ref{conf}) can be written in the following standard Schwarzschild form
\be\label{sc}
ds^2=-f(x)dt^2+\frac{dx^2}{f(x)}\ .
\ee
We find $f(x)=\frac{x^2}{\ell^2}\pm 2\mu\ x-1$, where $x=\frac{\sqrt{q}\ell^2}{\tanh(z\sqrt{q})}\pm\ell^2\mu$ and $\mu\equiv\sqrt{q-\frac{1}{\ell^2}}$, implying in turn that $q\geq\frac{1}{\ell^2}$. 

The metric (\ref{sc}) can also be extended to the whole space of $x$, in particular in the region in which $f(x)<0$ (inner horizon). 
Additionally, the metric (\ref{sc}) represents a BH geometry only in the case in which a boundary in $x=0$ is inserted. This is possible by exploiting the double definition sign choice of $\mu$. In particular we will then consider 
\be\label{sc2}
f(x)=\frac{x^2}{\ell^2}- 2\mu\ |x|-1\ .
\ee
In this case, the mass $\mu>0$ is a physical mass inserted at $x=0$. 

The two metrics (\ref{sc2}) and (\ref{conf}), with $\rho$ real, are obviously equal only in the chart $f(x)>0$, limiting the range of $x>x_h$ where $x_h=\ell^2(\sqrt{q}+\mu)$ is the  ``horizon" of the metric (\ref{sc}) and it is reached by $z\rightarrow\infty$.

The extension of $f(x)$ for negative values corresponds to the the second choice of (\ref{sol1}). The interior of the BH is thus described by a complex $\rho$ (note the shift in the coordinate $z$)
\be\label{solin}
e^{2\rho_{\rm in}}=-\frac{q\ell^2}{\cosh^2[(z_0+|z|)\sqrt{q}]} \ ,
\ee
where now $|x|=\sqrt{q}\ell^2 \tanh((z_0+|z|)\sqrt{q})+\ell^2\mu\leq x_h$ and $z_0=-\frac{1}{\sqrt{q}}\tanh^{-1}\left(\frac{\mu}{\sqrt{q}}\right)$ corresponds to $x=0$, i.e. the center of the BH. The two metrics in \eqref{sol1} are matched at $z\rightarrow\infty$. 

The complex Liouville field (\ref{solin}) can nevertheless be mapped into a real solution of another Liouville theory by shifting $\rho_{\rm in}=\tilde\rho_{\rm in}+i \frac{\pi}{2}$ so that
\be
e^{2\tilde\rho_{\rm in}}=\frac{q\ell^2}{\cosh^2[(z_0+|z|)\sqrt{q}]} \ .
\ee
Moreover, with this complex rotation, $\tilde\rho_{\rm in}$ is a real solution of 
\be\label{Lin}
\tilde\square\tilde\rho_{\rm in}=-\frac{1}{\ell^2}\ e^{2\tilde\rho_{\rm in}}+2\mu\delta(z)\ ,
\ee
where the boundary is, as it should, proportional to the BH mass $\mu$.
 
One can easily see that $\tilde \rho_{\rm in}$ and $\rho_{\rm out}$ are glued together at the point at infinity $\rho_{\rm in,out}\rightarrow -\infty$. Although,  (\ref{Lin}) come from an analytical rotation of the Liouville field $\rho$ in the complex plane, quantum mechanically, 
(\ref{Lout}) and (\ref{Lin}) represent two different theories called, respectively, spacelike and timelike Liouville theories \cite{witten}. 

The action describing \eqref{Lin} is (now and in the rest of the paper we drop the tildes)
\be
S_{c}=\int dzd\tau \left[\partial_\mu\rho_{\rm in}\partial^\mu\rho_{\rm in}-\frac{1}{\ell^2}e^{2\rho_{\rm in}}+2\mu\delta(z)\rho_{\rm in}\right]\ .
\ee 
Before going to the next section we will define the Euclidean coordinates used in the quantization procedure. 

First of all, we will Wick rotate time $t\rightarrow i\tau$ where the periodicity $\tau\equiv\tau+2\pi$ is imposed. It is then useful to use the complex coordinate $\zeta=2(z+i\tau)$ so that the auxiliary metric reads $d\tilde s^2=d\zeta d\bar\zeta$ and the interior and exterior Liouville equations read
\be
\partial_\zeta\partial_{\bar \zeta}\rho_{\rm in}&=&-\frac{4}{\ell^2}\ e^{2\rho_{\rm in}}+8\mu\delta(\zeta+\bar\zeta)\ ,\cr
\partial_\zeta\partial_{\bar \zeta}\rho_{\rm out}&=&+\frac{4}{\ell^2}\ e^{2\rho_{\rm out}}\ .
\ee

\section{Liouville theory description of black holes}

In Euclidean coordinates $(\zeta,\bar\zeta)$, the system (\ref{sys}) may be recast into the Liouville theory \cite{poly} (see for example \cite{ginsparg} for a review)
\be\label{Q}
S_L^{\pm}=\pm\frac{1}{4\pi}\int_\Gamma d^2\zeta\sqrt{\tilde g}\left[\partial_\mu\bar\rho_{\pm}\partial^\mu\bar\rho_{\pm}+Q_{\pm}\tilde R\bar\rho_\pm\pm 4\pi\lambda e^{2b\bar\rho_\pm}\right]\pm\int_{\partial\Gamma}Q_{\pm}\frac{K}{2\pi}\bar\rho_{\pm}d\theta\ ,
\ee
where $Q_{\pm}\equiv (b^{-1}\pm b)$. The choice $\pm$ respectively describes a spacelike and a timelike Liouville theory \cite{witten}, $\Gamma$ is the region where the theory lives with boundary $\partial\Gamma$ and $b$ is a coupling constant parameterizing the ``quantumness" of the system. Finally, $K$ is the extrinsic curvature and $\theta$ is the adapted coordinate to the boundary.

The Euclidean metric $\tilde g_{\alpha\beta}$ is an auxiliary metric that can be set to be flat (up to boundaries) after variation. 

The central charge of the theory (\ref{Q}), related with the semiclassical $\gamma$, is $c=1\pm 6 Q^2$. The factor $e^{2\frac{\bar\rho}{Q}}$ is the (quantum) conformal factor of the physical metric $g_{\alpha\beta}$. 

The theory (\ref{Q}) is invariant, up to a $c$-number anomaly, under the following conformal transformations
\be\label{conf2}
\zeta'&=&w(\zeta)\cr
\bar\rho'(\zeta',\bar\zeta')&=&\bar\rho_{\pm}(\zeta,\bar\zeta)-\frac{Q_{\pm}}{2}\ln\Big|\frac{\partial w}{\partial \zeta}\Big|^2\ .
\ee

The semiclassical equations (\ref{sys}) are obtained in the limit $b\rightarrow 0$ with $\rho=b\bar\rho\neq 0$ and $\lim_{b \to 0}4\pi\lambda b^2=\frac{1}{\ell^2}$. 

Note that this limit is by no means classical as the system is not weakly coupled here. The classical limit is indeed only obtained for $b\rightarrow 0$ but, conversely from before, $\lambda b\rightarrow 0$ and $b\bar\rho\rightarrow 0$. This is the reason why the limit $b\rightarrow 0$ is called {\it semiclassical}.
 
The boundary for the interior (the exterior has no boundary) is matched for
\be
\lim_{b\rightarrow 0}b\ Q_- K=\lim_{b\rightarrow 0}K=\mu\ ,
\ee
as expected the BH mass corresponds to the extrinsic curvature of the boundary.

We will now consider the quantization of (\ref{Q}) via the Euclidean path integral
\be\label{Z}
Z=\int{\cal D}\bar\rho e^{-S_L}\ .
\ee
The primary operators (or fields) of the conformal field theory (\ref{Q}) are $V_{\alpha}\equiv e^{2\alpha\bar\rho}$  where $\alpha$ are called the Liouville momenta \cite{17}. The theory can then be characterized by studying the n-points correlation functions (``measures") $\langle V_{\alpha_1}\ldots V_{\alpha_n}\rangle$. 

The operators $V_{\alpha}$ are labelled as ``heavy" (or semiclassical) if $\alpha_h=\frac{\eta}{b}$, where $\eta$ is kept constant when $b\rightarrow 0$. Heavy operators modify the saddle points of the path integral (\ref{Z}) and correspond to ``semiclassical" observables. In other words, in the semiclassical limit, an insertion of a heavy operator at the point $\zeta=\zeta_i$ would locally modify the Liouville equation by introducing a boundary term
\be\label{heavy}
 \square\rho_{\pm}=\pm 2\pi\lambda b^2\ e^{2\rho_{\pm}}-4\pi\eta\delta^2(\zeta-\zeta_i)\ .
\ee
Finally, ``light" operators are defined by the $\alpha_l=\sigma b$ where, again, $\sigma$ is kept fixed in the $b\rightarrow 0$ limit. Light operators do not modify the saddle points of (\ref{Z}) and thus they can be thought of as purely ``quantum". 

In this paper we will only consider heavy operators as we only wish to characterize a BH semiclassically, i.e. by defining it in terms of saddle points of (\ref{Z}). The study of light operators is postponed for a future work and it is probably related with the thermodynamical properties of the BH, which are not fully exploited in this work.

\subsection{A semiclassical black hole is a system with infinite central charge}

In the previous section, we showed that the semiclassical BH is described by two regions delimited by an horizon. The outer region is described by a spacelike Liouville theory, whereas, the inner region by a timelike Liouville theory. The two regions are then connected at the point at ``infinity" (the horizon). 

The key assumption of this work is that not only at the semiclassical level, but at the full quantum level, a two-dimensional BH is defined by "gluing" together a time-like and a space-like Liouville theory. Later on we shall give the exact meaning of "gluing together" these two theories at the full quantum level.

We will be only interested in the semiclassical limit $b\rightarrow 0$ of (\ref{Q}). The full quantum description of this limit has been exhaustively studied in \cite{witten}, so we will use the results of \cite{witten} without re-deriving them here.

In order for the Liouville field to be non singular at infinity with respect to the conformal transformation (\ref{conf2}) one needs that
\be
\bar\rho_\pm\sim -2 Q_{\pm}\ln r\ ,\ {\rm for}\ |\zeta|\equiv r\rightarrow\infty\ .
\ee
Thus, in order to match the two Liouville theories in the semiclassical limit at spatial infinity, i.e.
\be \label{match4}
\lim_{r\rightarrow\infty}\bar\rho_-=\lim_{r\rightarrow\infty}\bar\rho_+\ ,
\ee
one needs 
\be
Q_-\approx Q_+\ ,
\ee
which is exactly what happens in the $b\rightarrow 0$ limit, where $Q_{\pm}\rightarrow\infty$, or in the limit of an infinite number of degrees of freedom. 

Of course, as it happens in four-dimensions, the full quantum description of a BH replaces the BH horizon, which is a surface located at infinite physical distance from any point in the outer region, to a ``quantum" surface at some finite distance. This would in turn requires that 
\be
Q_-= Q_++{\cal O}(b)\ ,
\ee
i.e. the BH would not be constructed out of an infinite amount of states, but large. We can have a feeling about the perturbation of a finite $b$ in the following way: the central charge of a CFT counts the number of degrees of freedom $N$ of the system. As discussed before (for $b\rightarrow 0$) $c\sim Q^2\sim b^{-2}$. Thus in the small $b$ limit $b\sim N^{-1/2}$. We then found that
\be
Q_-= Q_++{\cal O}(\frac{1}{\sqrt{N}})\ ,
\ee
which is the typical correction related to BH thermodynamics in the exterior \cite{dvali}.

It is interesting to consider the Cardy entropy of the system. This is defined as $S=\sqrt{c E}$ where $E$ is a generic energy eigenvalue of the Liouville field theory. A standard result is that $E>{\cal O}(1)\times c$ \cite{cesar}, thus $S>{\cal O}(1)\times c$. Therefore, for a semiclassical BH (where $c_+\rightarrow \infty$) $E\sim c$ and
\be
S_{BH}\sim c_+\rightarrow\infty\ ,
\ee
where $S_{BH}$ is the BH entropy. The fact that a two-dimensional semiclassical BH (described by some CFT) has an entropy proportional to the central charge ($S_{BH}\sim c_{CFT}$), has been also found by \cite{solo} in a different set-up.

A couple of notes are in order. It would seem strange that the matching condition of the interior with the exterior of the BH does not involve the mass $\mu$, but only the global charges $Q_{\pm}$. However, conversely to the four-dimensional case, two-dimensional gravity does not have propagating gravitons thus, locally, the geometry is $AdS_2$ and therefore, geometrically, only ``global" matchings are required. Nevertheless, in the quantum case $b>0$, the matching of the extrinsic curvature with the mass would not follow the semiclassical identity $K=\mu$, thus deforming the interior and exterior geometries. In other words, the quantum match of the two Liouville theories would then depend on $\mu$. This dependence is presumably related to the Hawking flux across the horizon as shown, holographically, in \cite{procopio}. We leave however the full discussion of this quantum phase for future work.

\subsection{Saddle points of the quantum black hole}

We will now attempt to semiclassically characterize the two dimensional BH discussed before. In order to do that, we will ``probe" it by studying correlation functions of semiclassical (heavy) operators inside the horizon. To make our point we will only need to consider two heavy operators.

Up to three heavy operators, the general correlation function (the DOZZ formula\footnote{This stands for Dorn, Otto, Zamolodchikov, Zamolodchikov.}) has been found in \cite{dozz}. In the case of two heavy operators we have 
\be\label{dozz}
\langle V_{\eta}(\zeta_1,\bar \zeta_1)V_{\eta}(\zeta_2,\bar \zeta_2)\rangle=\delta(0)\frac{2\pi}{b}\frac{G(\eta)}{|\zeta_{12}|^{(1-2\eta)/b^2}} ,
\ee 
where $\zeta_{ij}\equiv \zeta_i-\zeta_j$. In the limit $b\rightarrow 0$ one finds, for the time-like Liouville theory \cite{witten}
\be\label{dozz}
\!\!\!\!\!\!\!\!\!\!\!\!\!\!\!\!\!G(\eta)\approx \left(e^{2\pi i\frac{(1-2\eta)}{b^2}}-1\right)\exp\{\frac{1}{b^2}\left[(1-2\eta)\ln(4\ell^2)+2\left((1-2\eta)\ln(1-2\eta)-(1-2\eta)\right)\right]\} .
\ee
In the semiclassical limit, the path integral \eqref{Z} is expected to be dominated by the saddle points of the action \cite{witten}. In other words, one would expect
\be\label{G}
G(\eta)= \sum_n\exp\left(-A(\rho_n)\right)\ ,
\ee
where the Liouville action $A$, after the insertion of heavy operators, is\footnote{Note that the integral \eqref{A} must be regularized for the matching \eqref{G}. We invite the interested reader to consult \cite{witten} for details.}
\be\label{A}
b^2 A=\int d^2x \left[S_L^--\frac{\eta}{\pi}\left(\delta^2(\zeta-\zeta_1)+\delta^2(\zeta-\zeta_2)\right)\rho\right]
\ee
and finally $\rho_n$ correspond to the dominant saddle points of the action $A$.

Note however, that these saddle points might not only be related to the real (``classical") solutions of the equation of motion for the Liouville field. Indeed, in the timelike Liouville theory, it turns out that some complex saddle points (i.e. complex solutions of the Liouville field equation) contribute equally to the classical saddle point in the path integral generating the DOZZ formula \eqref{dozz}. This has been shown in \cite{witten} and so while using the results of this paper we invite the interested reader to see the details there.

The insertion of a heavy operator at a point $\zeta=\zeta_i$ in the Liouville theory, is equivalent to insert a source term at the same point into the classical equation of motion (c.f. \eqref{heavy}). Around this source, the Liouville field behaves like\footnote{Note that our $\rho$ corresponds to $\phi_c/2$ in \cite{witten}.}
\be
\rho=-8\eta_i \ln|\zeta-\zeta_i|+{\cal O}(1)\ \ \ {\rm as}\ \ \ \zeta\rightarrow \zeta_i\ .
\ee
Having this in mind, it is relatively straightforward to find the complex solutions $\rho_n$ of the timelike version of \eqref{heavy}, once two heavy operators are inserted. These are \cite{witten}
\be\label{rs}
\!\!\!\!\!\!\!\!\!\!\!\!\!\rho_n=4\pi i\ n+2\ln(4\ell^2)-4\ln\kappa-4\ln\left(|\zeta-\zeta_1|^{2\eta}|\zeta-\zeta_2|^{2\eta}+\frac{|\zeta-\zeta_1|^{2-2\eta}|\zeta-\zeta_2|^{2-2\eta}}{\kappa^2(1-2\eta)^2|\zeta_{12}|^2}\right),
\ee
where $\kappa$ is an arbitrary complex number that, after integration in \eqref{G} will disappear. Thus, the path integral integration over $\kappa$ would generate a $\delta(0)$ as in \eqref{dozz}. Finally $n$ is an integer labeling the different complex solutions.

Substituting \eqref{rs} into \eqref{A}, one finds that the DOZZ formula \eqref{dozz} is matched as in \eqref{G} by a sum over the $n=0$ and $n=1$ saddle points of the timelike Liouville theory. In other words, the BH path integral is dominated by two distinct saddle points. It turns out that the same happens for three-point correlation functions of heavy operators, while for light operators the structure of saddle points is somehow more complicated \cite{witten}. 

{\it A clarification}: The two-dimensional semiclassical "Black Hole" studied here is defined as the system matching two CFTs, a timelike and a spacelike Liouville theory. Thus, the above calculation of the n-point correlation functions, corresponds to the n-point correlation functions behind the horizon. 
Of course, the results used above are only valid in the limit $c\rightarrow \infty$ which, in comparison with the four-dimensional case, corresponds to an infinitely large BH. Therefore, the BH in this limit is not globally distorted by the insertion of heavy operators (although locally it is). 

\section{Interpretation}

As discussed in the introduction, the manifestation of the information loss paradox for BHs is  that correlation functions of matter fields are {\it not} dominated by the classical BH solution in the path integral, i.e., even in the semiclassical limit $\hbar\rightarrow 0$ 
\be
\langle O_1 O_2\rangle\neq \langle O_1 O_2\rangle_{BH}\ .
\ee
One thus expects that other (quantum) configurations should contribute to restore information, i.e.
\be\label{sumsaddle}
\langle O_1 O_2\rangle= \langle O_1 O_2\rangle_{BH}+\sum_{g} \langle O_1 O_2\rangle_{g}\ ,
\ee
where $g$ are some other ''gravity" saddle points with large entropy but same boundary conditions as in the BH configuration. The requirement of large entropy is essential to avoid exponential suppressions of the $g$ contributions in \eqref{sumsaddle}. In other words, the $g$ configurations must be macroscopic. 

A simple possibility is that a BH is a condensate with co-existing degenerate quantum states, i.e. a condensate at a quantum critical point, as suggested in \cite{dvali}. We will show that in our two-dimensional case this is indeed the most natural interpretation.

First of all, a CFT with negative central charge, as in the time-like Liouville theory, is non-unitary. The only physical explanation of this fact is that the Liouville field, describing the interior of the BH, cannot be fundamental. 

The c-theorem \cite{c-theorem} states that a flow of unitary CFTs can only satisfy the inequality
\be
c^{\rm UV}\geq c^{\rm IR}\ ,
\ee
in other words the number of degrees of freedom of the system must always increase at higher and higher energies, if the theory is Wilsonian.

The c-theorem, extended to non-unitary CFT, would imply that, for the effective central charges
\be
c_{\rm eff}=c-24\Delta_{\rm min}\ ,
\ee
where $\Delta_{\rm min}$ is the lowest conformal weight in the theory, 
\be
c_{\rm eff}^{\rm UV}\geq c_{\rm eff}^{\rm IR}\ .
\ee
Unfortunately, generically this extension is false \cite{ref}.

However, in the case in which the IR CFT describe fluctuations on a Bose-Einstein condensate of the UV CFT, this extension is most probably correct. In this case indeed, the number of the effective degrees of freedom of the condensate can only be less than of the microscopic degrees of freedom forming the condensate. Let us apply this discussion to the BH.

The exterior of the BH is a unitary CFT and thus $\Delta_{\rm min}^{+}=0$ corresponding to the conformal vacuum. The interior has instead negative conformal weights. The conformal weights in the timelike Liouville theory are $\Delta=\alpha(Q_--\alpha)$ \cite{witten} and thus we have that $c_{\rm eff}^-=13+$"light" operators contribution (in the $b\rightarrow 0$ limit). This implies that
\be
c_{\rm eff}^{\rm +}\gg c_{\rm eff}^{\rm -}\ ,\ {\rm for}\ ,\ b\rightarrow 0\ .
\ee
As we match the two CFTs at the horizon, we see that the exterior CFT must be the UV part of the "extended" c-theorem. This seemly paradoxical result can only be accepted, as discussed above, if the interior CFT is an effective description of some collective excitations on a system with a non-trivial background. While the exterior CFT genuinely describe a fundamental theory.

One may even reverse the argument. Since $c_{\rm eff}^{\rm +}\gg c_{\rm eff}^{\rm -}$ and since the two CFTs {\it must} flow one to the other, the non-unitary Liouville theory can only describe here fluctuations of a condensate of the unitary theory.

There is indeed a precise mapping of a two-dimensional CFT with large negative charge (as in our case) and two two-dimensional interacting quasi-condensates defined by two degenerate quantum states \cite{science}. The CFT correlation functions are then mapped to the distribution function of the fringe amplitudes due to the interference of the two condensates. In \cite{science} it has been shown that the parameter $b$ of our CFT is related to the Luttinger parameter ${\cal K}$ ($b=(2{\cal K})^{-1}$), describing the interaction strength of the two condensates. In particular, large $\cal K$ implies weak coupling, which match our definition of semi-classicality.

\subsection{Two dimensional black hole in AdS/CFT}

There is an intriguing similarity with the AdS/CFT analysis of \cite{procopio} that we like to point out here. 

In the Randall-Sundrum set-up \cite{RS} where ``the Universe" is a $(D-1)$-brane embedded in an $AdS_D$ spacetime, it has been shown, for the $D=4$ case, that a static brane BH do not exist \cite{bruni}. A BH solution might however only exist if extra-energy is supplemented in form of a non-vanishing brane Ricci scalar. Interestingly, this supplemental energy, for a given asymptotic mass, exactly matches the trace anomaly of a quantum BH. In \cite{emparan} (and then in \cite{casadio}) it has then put forward the idea that brane BH are holographically dual to quantum (semiclassical) BHs.  

In two dimensions, this duality has been checked in \cite{procopio} where a string (D1-brane) configuration, reproducing \eqref{sc} in a BTZ \cite{btz} bulk, has been constructed (note that, as discussed before, the metric \eqref{sc} is semiclassical) \footnote{As an extra non-trivial check of the duality, it was then shown that the thermodynamical properties of the BH solution \eqref{sc} are, using the AdS/CFT dictionary of \cite{maldacena}, matched with the thermodynamical properties of the bulk BTZ BH.}.

In other words, in \cite{procopio}, the semiclassical Black Hole described in this paper, has been shown to be holographically dual to the gravitational system produced gluing together, along two intersecting strings, two copies of a BTZ BH. 

The physics of the two intersecting strings, identifying the boundary CFTs, is then dual to the physics of our two-dimensional quantum Black Hole, in the spirit of AdS/CFT \cite{maldacena}. Thus, it seems natural to state that the {\it two} saddle points needed to study quantum mechanically the Liouville theory are dual boundary descriptions of the gravitational background described by {\it two} intersecting strings in a BTZ bulk.

One would be in principle puzzled by the fact that the boundary theory is non-unitary. This is however not new in holographic correspondences inside an horizon, e.g. in dS/CFT \cite{dscft} and, more generically, in FRW/CFT \cite{frwcft}. In particular, it has been argued in \cite{frw2} that the CFT in the FRW/CFT is not UV completable. This of course very well match with the idea that the CFT describing the interior of an horizon is only an effective theory of quantum fluctuations on a condensate. From the gravity point of view the lost of unitarity is obviously related to the use of a classical saddle point: the BTZ black hole.

\subsection{Absence of firewalls in $b>0$}

It has been argued in \cite{firewall} that an observer falling into the BH should see a ''firewall" crossing the horizon. In other words the horizon is not a smooth surface as in GR. This result has been obtained by considering the following three main assumptions
\begin{itemize}
\item There is an horizon,
\item The theory on a BH background is unitary,
\item The BH has a non-vanishing temperature.
\end{itemize}
In the $b\rightarrow 0$ limit we do expect a firewall as a free falling observer, hitting the horizon, will experience a sudden change in the Liouville theory (from spacelike to timelike). However, we believe this is only an artifact of the   $b\rightarrow 0$ limit.

In the $b>0$ case, as discussed before, we expect the ''horizon" to live at some finite point in space, dependent upon $\mu$ (the BH mass) and $b$. In this case, the central charges of the two CFTs, describing the interior and exterior of the BH, will no longer diverge ($|c_{\pm}|<\infty$). 

In the quantum ($b>0$) case we thus expect a transition region flowing the UV central charge (the exterior) to the IR one (the interior). This transition must involve a flux of degrees of freedom at the ''horizon" that should match the usual thermodynamical properties of quantum BHs with finite entropy and temperature, at least for large enough $|c_\pm|$. 

In this case, a free falling observer will no longer suddenly see a firewall but a smooth flow from the fundamental CFT describing the exterior to the condensate of the interior. Thus, no firewall may appear in $b>0$.

It is interesting to see this from a different point of view. In \cite{mark}\footnote{Note that, although in \cite{mark} the author still consider two interacting CFTs, his system is physically different from ours. The two CFTs in \cite{mark}, as in \cite{maldasuss}, correspond to the two outer horizon regions of the analytical extension of an AdS BH. In our case instead the two CFTs live, respectively, inside and outside the BH horizon.} it has been argued that, although an asymptotically large AdS BH in thermal equilibrium probably has a firewall at the horizon, once this system is forced to start an evaporation (as presumably would happen beyond the semiclassical regime), the firewall disappears. This conclusion nicely matches our system if, for $c_\pm={\cal O}(1)$, the BH would no longer be in thermal equilibrium. 

One can in fact imagine to have a large BH ($b>0$ but still small) with initially large central charge $c_+$, where a free falling observer would experience a hard wall at the horizon. This hard wall, as described before, is related to the rapid change from a spacelike to a timelike Liouville theory. This rapid change, will indeed make a free-falling observer experience a sudden change in the correlation functions of any operator at his/her hand. However, in the $b>0$ we expect a smooth flow of degrees of freedom between the interior and the exterior of the BH (i.e. the BH evaporates). In this case, the thermal equilibrium would be broken and at the same time, a free-falling observer would see a smooth transition between the central charges of the two Liouville theories as explained before. In other words, again, the appearance of a firewall would just be a coarse grained effect due to the $b\rightarrow 0$ limit, or in the language of \cite{dvali} $N\rightarrow \infty$, limit.

\section{Conclusions} 

In this paper we have argued that a two dimensional (quantum) BH is well described by matching two CFTs at the ''horizon". In particular, in the semiclassical limit, where the horizon is at ``infinity" the two theories are a spacelike (BH exterior) and a timelike (BH interior) Liouville field theories.

The BH interior is characterized by two degenerate saddle points of the Liouville theory. 
This has an immediate correspondence in AdS/CFT as the configuration obtained by intersecting two strings in a three-dimensional BTZ BH.

As the timelike Liouville theory is not unitary, physically, it can only describe some collective excitations on a non-trivial background. This is the typical case of a condensate and indeed, following \cite{science}, we mapped our BH interior to a system of interacting one-dimensional Bose liquids. 

We have not answered the question of what the condensate is, or more concretely, how does the condensate dynamically form from the spacelike Liouville field theory, and we leave this important question for future work. 

Finally, we have argued that, although in the semiclassical limit a free falling observer should see a firewall passing through the horizon, this would not happen in the $b>0$ (finite entropy) case.

\section*{Acknowledgements}
I would like to thank: Daniel Flassig, Cesar Gomez and Alex Pritzel for helping me to acquire a deeper understanding of quantum black holes and Liouville theory. Jose Barbon for commenting on the first draft and on free-falling observers. Davide Fioravanti and Francesco Ravanini for clarifications on CFTs. Daniele Oriti for discussions on unitary theories. Federico Piazza for clarifications about the Bose-Einstein condensate of \cite{science}. Christophe Galfard for discussions on the gravity path integral and its relation to the information paradox. Daniel Harlow for correspondences on the timelike Liouville theory. I am supported by the Humboldt foundation.


\begin{thebibliography}{999}
\bibitem{hoop}
K.~S.~Thorne, in {\it Magic without magic: John Archibald Wheeler}, edited
by J. Klauder (Freeman, S. Francisco) 1972, p. 231.
\bibitem{sarah}
G.~Dvali, S.~Folkerts and C.~Germani,
  %``Physics of Trans-Planckian Gravity,''
  Phys.\ Rev.\ D {\bf 84} (2011) 024039
  [arXiv:1006.0984 [hep-th]].
  %%CITATION = ARXIV:1006.0984;%%
  %35 citations counted in INSPIRE as of 04 Jul 2013
  \bibitem{nohair}
  C.~W.~Misner, K.~S.~Thorne and J.~A.~Wheeler,
  ``Gravitation,''
  San Francisco 1973, 1279p
  %8 citations counted in INSPIRE as of 04 Jul 2013
  \bibitem{how}
  S.~W.~Hawking,
  %``Black hole explosions,''
  Nature {\bf 248} (1974) 30.
  %%CITATION = NATUA,248,30;%%
  %1628 citations counted in INSPIRE as of 04 Jul 2013
\bibitem{barbon}
J.~L.~F.~Barbon and E.~Rabinovici,
  %``Very long time scales and black hole thermal equilibrium,''
  JHEP {\bf 0311} (2003) 047
  [hep-th/0308063].
  %%CITATION = HEP-TH/0308063;%%
  %42 citations counted in INSPIRE as of 04 Jul 2013
\bibitem{malda}
J.~M.~Maldacena,
  %``Eternal black holes in anti-de Sitter,''
  JHEP {\bf 0304} (2003) 021
  [hep-th/0106112].
  %%CITATION = HEP-TH/0106112;%%
  %296 citations counted in INSPIRE as of 04 Jul 2013
\bibitem{lasthaw}
S.~W.~Hawking,
  %``Information loss in black holes,''
  Phys.\ Rev.\ D {\bf 72} (2005) 084013
  [hep-th/0507171].
  %%CITATION = HEP-TH/0507171;%%
  %171 citations counted in INSPIRE as of 04 Jul 2013
\bibitem{dvali}
G.~Dvali and C.~Gomez,
  %``black hole's Quantum N-Portrait,''
  Fortsch.\ Phys.\  {\bf 61} (2013) 742
  [arXiv:1112.3359 [hep-th]];
  %%CITATION = ARXIV:1112.3359;%%
  %21 citations counted in INSPIRE as of 04 Jul 2013
G.~Dvali and C.~Gomez,
  %``black hole's 1/N Hair,''
  Phys.\ Lett.\ B {\bf 719} (2013) 419
  [arXiv:1203.6575 [hep-th]].
  %%CITATION = ARXIV:1203.6575;%%
  %15 citations counted in INSPIRE as of 04 Jul 2013
G.~Dvali and C.~Gomez,
  %``black holes as Critical Point of Quantum Phase Transition,''
  arXiv:1207.4059 [hep-th];
  %%CITATION = ARXIV:1207.4059;%%
  %9 citations counted in INSPIRE as of 04 Jul 2013
G.~Dvali and C.~Gomez,
  %``black hole Macro-Quantumness,''
  arXiv:1212.0765 [hep-th].
  %%CITATION = ARXIV:1212.0765;%%
  %5 citations counted in INSPIRE as of 04 Jul 2013
  \bibitem{alex}
  D.~Flassig, A.~Pritzel and N.~Wintergerst,
  %``black holes and Quantumness on Macroscopic Scales,''
  arXiv:1212.3344 [hep-th].
  %%CITATION = ARXIV:1212.3344;%%
  %2 citations counted in INSPIRE as of 04 Jul 2013

\bibitem{bek}
J.~D.~Bekenstein,
  %``Black holes and entropy,''
  Phys.\ Rev.\ D {\bf 7} (1973) 2333.
  %%CITATION = PHRVA,D7,2333;%%
  %2389 citations counted in INSPIRE as of 04 Jul 2013
\bibitem{norma}
N.~G.~Sanchez,
  %``Semiclassical Quantum Gravity And Liouville Theory: A Complete Solution To The Back Reaction Problem In Two-dimensions,''
  Nucl.\ Phys.\ B {\bf 266} (1986) 487.
  %%CITATION = NUPHA,B266,487;%%
  %28 citations counted in INSPIRE as of 04 Jul 2013
\bibitem{birrel}
N.~D.~Birrell and P.~C.~W.~Davies,
  ``Quantum Fields In Curved Space,''
  Cambridge, Uk: Univ. Pr. ( 1982) 340p
  \bibitem{kallosh}
  R.~Gastmans, R.~Kallosh and C.~Truffin,
  %``Quantum Gravity Near Two-Dimensions,''
  Nucl.\ Phys.\ B {\bf 133} (1978) 417.
  %%CITATION = NUPHA,B133,417;%%
  %108 citations counted in INSPIRE as of 04 Jul 2013
\bibitem{procopio}
C.~Germani and G.~P.~Procopio,
  %``Two-dimensional Quantum black holes, Branes in BTZ and Holography,''
  Phys.\ Rev.\ D {\bf 74} (2006) 044012
  [hep-th/0605068].
  %%CITATION = HEP-TH/0605068;%%
  %1 citations counted in INSPIRE as of 04 Jul 2013
\bibitem{hartle}
J.~B.~Hartle and S.~W.~Hawking,
  %``Wave Function of the Universe,''
  Phys.\ Rev.\ D {\bf 28} (1983) 2960.
  %%CITATION = PHRVA,D28,2960;%%
  %1682 citations counted in INSPIRE as of 04 Jul 2013
\bibitem{gibbons}
G.~W.~Gibbons and S.~W.~Hawking,
  %``Cosmological Event Horizons, Thermodynamics, and Particle Creation,''
  Phys.\ Rev.\ D {\bf 15} (1977) 2738.
  %%CITATION = PHRVA,D15,2738;%%
  %1366 citations counted in INSPIRE as of 04 Jul 2013
\bibitem{witten}
  D.~Harlow, J.~Maltz and E.~Witten,
  %``Analytic Continuation of Liouville Theory,''
  JHEP {\bf 1112} (2011) 071
  [arXiv:1108.4417 [hep-th]].
  %%CITATION = ARXIV:1108.4417;%%
  %18 citations counted in INSPIRE as of 04 Jul 2013
\bibitem{poly}
A.~M.~Polyakov,
  %``Quantum Geometry of Bosonic Strings,''
  Phys.\ Lett.\ B {\bf 103} (1981) 207.
  %%CITATION = PHLTA,B103,207;%%
  %2202 citations counted in INSPIRE as of 04 Jul 2013
\bibitem{ginsparg}
P.~H.~Ginsparg and G.~W.~Moore,
  ``Lectures on 2-D gravity and 2-D string theory,''
  In *Boulder 1992, Proceedings, Recent directions in particle theory* 277-469. and Yale Univ. New Haven - YCTP-P23-92 (92,rec.Apr.93) 197 p. and Los Alamos Nat. Lab. - LA-UR-92-3479 (92,rec.Apr.93) 197 p
  [hep-th/9304011].
  %%CITATION = HEP-TH/9304011;%%
  %369 citations counted in INSPIRE as of 04 Jul 2013
\bibitem{17}
A.~A.~Belavin, A.~M.~Polyakov and A.~B.~Zamolodchikov,
  %``Infinite Conformal Symmetry in Two-Dimensional Quantum Field Theory,''
  Nucl.\ Phys.\ B {\bf 241} (1984) 333.
  %%CITATION = NUPHA,B241,333;%%
  %2695 citations counted in INSPIRE as of 04 Jul 2013
  \bibitem{cesar}
  L. Alvarez-Gaume and C. Gomez, ``Topics in Liouville theory,''\\
  In *Trieste 1991, Proceedings, String theory and quantum gravity '91* 142-177 and CERN Geneva - TH. 6175 (91/07,rec.Sep.) 35 p

  \bibitem{solo}
  S.~N.~Solodukhin,
  %``Conformal description of horizon's states,''
  Phys.\ Lett.\ B {\bf 454} (1999) 213
  [hep-th/9812056].

\bibitem{dozz}
H.~Dorn and H.~J.~Otto,
  %``Two and three point functions in Liouville theory,''
  Nucl.\ Phys.\ B {\bf 429} (1994) 375
  [hep-th/9403141];
  %%CITATION = HEP-TH/9403141;%%
  %211 citations counted in INSPIRE as of 04 Jul 2013
A.~B.~Zamolodchikov and A.~B.~Zamolodchikov,
  %``Structure constants and conformal bootstrap in Liouville field theory,''
  Nucl.\ Phys.\ B {\bf 477} (1996) 577
  [hep-th/9506136].
  %%CITATION = HEP-TH/9506136;%%
  %351 citations counted in INSPIRE as of 04 Jul 2013
  \bibitem{c-theorem}
   A.~B.~Zamolodchikov,
  %``Irreversibility of the Flux of the Renormalization Group in a 2D Field Theory,''
  JETP Lett.\  {\bf 43} (1986) 730
   [Pisma Zh.\ Eksp.\ Teor.\ Fiz.\  {\bf 43} (1986) 565].
  %%CITATION = JTPLA,43,730;%%
  %902 citations counted in INSPIRE as of 21 Jul 2013
   \bibitem{ref}
   C.~Dunning,
  %``Massless flows between minimal W models,''
  Phys.\ Lett.\ B {\bf 537} (2002) 297
  [hep-th/0204090].
  %%CITATION = HEP-TH/0204090;%%
  %4 citations counted in INSPIRE as of 27 Sep 2013
\bibitem{science}
V. Gritsev, E. Altman, E. Demler, and A. Polkovnikov,
%Full quantum distribution of contrast in interference experiments between interacting one-dimensional Bose liquids,
Nature Physics
{\bf 2}, 705 (2006).
\bibitem{dscft}
A.~Strominger,
  %``The dS / CFT correspondence,''
  JHEP {\bf 0110} (2001) 034
  [hep-th/0106113].
  %%CITATION = HEP-TH/0106113;%%
  %557 citations counted in INSPIRE as of 21 Jul 2013
\bibitem{frwcft}
Y.~Sekino and L.~Susskind,
  %``Census Taking in the Hat: FRW/CFT Duality,''
  Phys.\ Rev.\ D {\bf 80} (2009) 083531
  [arXiv:0908.3844 [hep-th]].
  %%CITATION = ARXIV:0908.3844;%%
  %27 citations counted in INSPIRE as of 21 Jul 2013
  \bibitem{frw2}
  D.~Harlow and L.~Susskind,
  %``Crunches, Hats, and a Conjecture,''
  arXiv:1012.5302 [hep-th].
  %%CITATION = ARXIV:1012.5302;%%
  %28 citations counted in INSPIRE as of 21 Jul 2013
\bibitem{RS}
L.~Randall and R.~Sundrum,
  %``An Alternative to compactification,''
  Phys.\ Rev.\ Lett.\  {\bf 83} (1999) 4690
  [hep-th/9906064].
  %%CITATION = HEP-TH/9906064;%%
  %4816 citations counted in INSPIRE as of 04 Jul 2013
\bibitem{bruni}
M.~Bruni, C.~Germani and R.~Maartens,
  %``Gravitational collapse on the brane,''
  Phys.\ Rev.\ Lett.\  {\bf 87} (2001) 231302
  [gr-qc/0108013].
  %%CITATION = GR-QC/0108013;%%
  %116 citations counted in INSPIRE as of 04 Jul 2013
\bibitem{emparan}
T.~Tanaka,
  %``Classical black hole evaporation in Randall-Sundrum infinite brane world,''
  Prog.\ Theor.\ Phys.\ Suppl.\  {\bf 148} (2003) 307
  [gr-qc/0203082];
  %%CITATION = GR-QC/0203082;%%
  %127 citations counted in INSPIRE as of 04 Jul 2013
R.~Emparan, A.~Fabbri and N.~Kaloper,
  %``Quantum black holes as holograms in AdS brane worlds,''
  JHEP {\bf 0208} (2002) 043
  [hep-th/0206155].
  %%CITATION = HEP-TH/0206155;%%
  %139 citations counted in INSPIRE as of 04 Jul 2013
\bibitem{casadio}
R.~Casadio and C.~Germani,
  %``Gravitational collapse and black hole evolution: Do holographic black holes eventually 'anti-evaporate'?,''
  Prog.\ Theor.\ Phys.\  {\bf 114} (2005) 23
  [hep-th/0407191];
  %%CITATION = HEP-TH/0407191;%%
  %34 citations counted in INSPIRE as of 04 Jul 2013
R.~Casadio and C.~Germani,
  %``Gravitational collapse and evolution of holographic black holes,''
  J.\ Phys.\ Conf.\ Ser.\  {\bf 33} (2006) 434
  [hep-th/0512202].
  %%CITATION = HEP-TH/0512202;%%
  %3 citations counted in INSPIRE as of 04 Jul 2013
  \bibitem{btz}
  M.~Banados, C.~Teitelboim and J.~Zanelli,
  %``The Black hole in three-dimensional space-time,''
  Phys.\ Rev.\ Lett.\  {\bf 69} (1992) 1849
  [hep-th/9204099].
  %%CITATION = HEP-TH/9204099;%%
  %1539 citations counted in INSPIRE as of 04 Jul 2013
\bibitem{maldacena}
J.~M.~Maldacena,
  %``The Large N limit of superconformal field theories and supergravity,''
  Adv.\ Theor.\ Math.\ Phys.\  {\bf 2} (1998) 231
  [hep-th/9711200].
  %%CITATION = HEP-TH/9711200;%%
  %9078 citations counted in INSPIRE as of 04 Jul 2013
  \bibitem{firewall}
  A.~Almheiri, D.~Marolf, J.~Polchinski and J.~Sully,
  %``Black Holes: Complementarity or Firewalls?,''
  JHEP {\bf 1302} (2013) 062
  [arXiv:1207.3123 [hep-th]]. cf. S. L. Braunstein, "Black hole entropy as entropy of entanglement, or it's curtains for the equivalence principle," [arXiv:0907.1190v1 [quant-ph]] published as S. L. Braunstein, S. Pirandola and K. Zyczkowski, "Better Late than Never: Information Retrieval from Black Holes," Physical Review Letters 110, 101301 (2013) for a similar prediction from different assumptions.
  %%CITATION = ARXIV:1207.3123;%%
  %72 citations counted in INSPIRE as of 21 Jul 2013
  \bibitem{mark}
  M.~Van Raamsdonk,
  %``Evaporating Firewalls,''
  arXiv:1307.1796 [hep-th].
  %%CITATION = ARXIV:1307.1796;%%
  %2 citations counted in INSPIRE as of 23 Jul 2013
\bibitem{maldasuss}
J.~Maldacena and L.~Susskind,
  %``Cool horizons for entangled black holes,''
  arXiv:1306.0533 [hep-th].
  %%CITATION = ARXIV:1306.0533;%%
  %10 citations counted in INSPIRE as of 23 Jul 2013

  \end{thebibliography}
\end{document}